\documentclass[prd,aps,letterpaper,floatfix,superscriptaddress,preprintnumbers,11pt,nofootinbib]{revtex4}
\usepackage{graphicx}  
\usepackage{dcolumn}   
\usepackage{bm}        
\usepackage{amssymb}   
\usepackage{amsmath}
\usepackage{slashed}
\usepackage{braket,slashed,bm}
\usepackage{array,multirow}
\usepackage[normalem]{ulem}
\usepackage{xcolor,cancel,youngtab}

\usepackage[T1]{fontenc} 

\usepackage{mathrsfs}

\usepackage{booktabs}
\usepackage{adjustbox}
\usepackage{mathtools}

\usepackage{soul}

\usepackage{tikz}
\usetikzlibrary{arrows,decorations.pathmorphing,backgrounds,positioning,fit,petri,automata,shadows,calendar,mindmap,
decorations.markings,calc,arrows.meta,bending}

\definecolor{labelkey}{rgb}{0,0.5,0.0}

\usepackage{xparse}
\usepackage{etoolbox}

\usepackage{feynmp}
\usepackage{graphicx,psfrag,amsmath,subfigure}
\usepackage{color}
\usepackage{slashed}
\usepackage{xcolor}
\usepackage{multirow}
\usepackage{ulem}
\usepackage{xcolor}
\usepackage{natbib}
\usepackage{notoccite}
\usepackage[colorlinks]{hyperref}
\hypersetup{ 
    colorlinks=true,
    linkcolor=blue,
    filecolor=magenta,      
    urlcolor=cyan,
}
\usepackage{hypcap}
\def\lsim{\mathrel{\raise.3ex\hbox{$<$\kern-.75em\lower1ex\hbox{$\sim$}}}}
\def\gsim{\mathrel{\raise.3ex\hbox{$>$\kern-.75em\lower1ex\hbox{$\sim$}}}}

\newcommand{\dis}{\displaystyle}
\newcommand{\be}{\begin{equation}} 
\newcommand{\ee}{\end{equation}} 
\newcommand{\bea}{\begin{eqnarray}}
\newcommand{\eea}{\end{eqnarray}}
\newcommand{\bee}{\begin{eqnarray*}}
	\newcommand{\eee}{\end{eqnarray*}}

\newcommand{\vev}[1] {\langle #1 \rangle}

\def\nn{\nonumber\\ }

\def\vev#1{\left\langle #1 \right\rangle}

\def\gcZ{{\overline g_{Z}}}

\def\sc{{\overline s}}

\def\q{\mathsf{q}}

\renewcommand{\O}{\mathcal{O}}
\newcommand{\op}[3]{\O^{#2,#3}_{#1}}

\newcommand{\nnn}{\nonumber\\[-0.4cm] }

\NewDocumentCommand{\Op}{ m m O{} o }{
	\O^{\ifblank{#3}{}{#3,}#2 }_{\IfNoValueTF{#4}{#1}{\substack{#1\\#4}}}
}
\NewDocumentCommand{\lwc}{ m m O{} o }{
	L^{\ifblank{#3}{}{#3,}#2 }_{\IfNoValueTF{#4}{#1}{\substack{#1\\#4}}}
}
\NewDocumentCommand{\dlwc}{ m m O{} o }{
	{\dot L}^{\ifblank{#3}{}{#3,}#2 }_{\IfNoValueTF{#4}{#1}{\substack{#1\\#4}}}
}

\begin{document}
\draft
\title{ Constraints on scalar $q^2 l^2$ operators from $\pi^0 \rightarrow \mu e$}
\title{ Bounds or Limits on  $q^2 l^2$ operators from $\pi^0 \rightarrow \mu e$}
 \title{ Restricting $q^2 l^2$ operators from $\pi^0 \rightarrow \mu e$}
  \author{Mathew Thomas Arun}
 \email{mathewthomas@iisertvm.ac.in}
 \affiliation{School of Physics, Indian Institute of Science Education and Research, Thiruvananthapuram, India}
 \author{Priyanka Lamba} 
 \email{priyanka.lamba@fuw.edu.pl}
 \affiliation{Institute of Theoretical Physics, Faculty of Physics,
University of Warsaw, Poland}
 \author{Sudhir K. Vempati}
 \email{vempati@iisc.ac.in}
 \affiliation{Centre for High Energy Physics, Indian Institute of Science, 
 Bangalore, India}

\vspace{0.8cm}
\begin{abstract}
In this paper we consider semileptonic lepton flavor violating operators of the type $q^2l^2$ in  low energy effective theory (LEFT). At the chiral scale, we match these operators to chiral perturbation theory ($\chi$PT) and place constraints from the process $\pi^0\rightarrow \mu^+ \,\, e^-$. These bounds are shown to depend on the chiral nature of the operators. The scalar operators
are significantly more constrained compared to the vector operators. We then compare the limits  from $\mu \to e$ conversion in Nuclei and show that the limits on scalar operators are within an order of magnitude of the corresponding limits from $\mu \to e$ conversion in Ti. On the other hand, the limits on vector operators are however much weaker. Towards the end, we evolve the LEFT operators to W-boson mass scale using RGE, and match them to the Standard Model effective field theory (SMEFT) operators. We, then, derive the constraints on the parameter space of Leptoquark models that could generate these SMEFT operators at tree level.

\end{abstract}

\maketitle

\section{Introduction}
\label{sec:intro}
Neutral pion decays of the type $ \pi^0 \to \mu e$ are excellent probes for studying lepton number and flavour violation. To our knowledge 
the first searches for this process constrains the Branching fraction of neutral pions decaying to the combination ($\mu^+ e^- + ~ \mu^- e^+$) to 
$1.72 \times 10^{-8}$ \cite{ParticleDataGroup:1998hll}. Whereas, later experiments performed at Brookhaven AGS, BNL E777\cite{PhysRevLett.64.2450}, BNL E865\cite{BNL:1998apv}, searching for the decays of $K_L^0 \rightarrow \mu^\pm e^\mp$ \cite{BNL:1998apv}, $K_L^0 \rightarrow \pi^0 \mu^\pm e^\mp$ \cite{Arisaka:1998uf,Appel:2000tc} and $K^+ \rightarrow \pi^+ \mu^+ e^-$ ($K_{\pi \mu e}$)\cite{Appel:2000wg}, in particular, the one looking for Lepton Flavour Violating $K^+ \rightarrow \pi^+ \ell \ell'$ decays, restricts the 
process $\pi^0 \rightarrow \mu e$ through the subprocess $K^+ \rightarrow \pi^+ \pi^0 (\rightarrow \mu e)$. The Brookhaven experiments, using them, place an upper limit on the Branching Fraction for the $\pi^0 \to \mu^+ e^-$ process as $BR(\pi^0 \to \mu^+ e^-)  < 3.8 \times 10^{-10}$ and $BR(\pi^0 \to \mu^- e^+)  ~< 3.4 \times 10^{-9}$. Later, the KTeV collaboration at the Fermilab has improved the limits on the combination 
to BR($\pi^0 \to \mu^\mp e^\pm$)$ < 3.59 \times 10^{-10}$ \cite{KTeV:2007cvy} and recently, the NA62 experiment at CERN constraint them further to BR($\pi^0 \to \mu^- e^+$)$ < 3.2 \times 10^{-10}$ \cite{NA62:2021zxl}. A summary of the present limits is presented
in Table \ref{expt1}.

\begin{table}[h]
\begin{tabular}{|c|c|c|}
\hline
Process & Experiment & Limit \\
\hline
$\pi^0 \to \mu^+ e^- $& E865 BNL & $3.4 \times 10^{-10}$ \cite{Appel:2000wg}\\ 
$\pi^0 \to \mu^- e^+ $& NA62 CERN&  $3.2 \times 10^{-10}$ \cite{NA62:2021zxl} \\
$\pi^0 \to \mu^\mp e^\pm$ & KTeV Fermilab & $3.6 \times 10^{-10}$ \cite{KTeV:2007cvy} \\
\hline
\end{tabular}
\caption{A list of the present limits on the Lepton number violating process
$\pi^0 \to \mu e$.}
\label{expt1}
\end{table}

It would be interesting to study the limits from these experiments on low energy effective operators. 
For this, one needs to match the effective operators to chiral perturbation theory. We do that in this article and derive the limits. To our knowledge this is first time this process is used to obtain the limits \footnote{While finishing this paper, Ref.\cite{Hoferichter:2022mna} appeared on the arxiv, addressing the same issue but in a way complementary to our work.}. 
In general, due to the presence of large hadronic uncertainties, pion decays (mostly charged ones) are more suitable for studying ratios like lepton flavor universality etc, where purely leptonic processes like $\mu \to e \gamma$ are good to study the lepton number/flavour violation in the leptonic sector. However, there exists other decay modes like $K_L \to \mu e $ which provide strong constraints on the effective theory operators. 

 Large flavor violation in the neutral lepton sector is very well established through neutrino oscillation phenomenology. The (large) neutrino transitions should typically lead to charged lepton flavor violation(cLFV). If neutrinos attain masses through  interactions with Higgs of the Standard Model, the resultant cLFV is understood to be very small, leading to the Branching Fraction of the most prominent channel $BR(\mu \rightarrow e \gamma) \sim 10^{-48}$. However, in the presence of new particles, with Weak scale masses, we expect the cLFV rates to be within the range of present and future upcoming experiments. The cLFV is a highly model dependent phenomena, such that, the predicted rates for various cLFV processes vary significantly from one UV model to another and with the  parameter space of the model. For example in supersymmetric models the rates can be large for supersymmetric particles between masses of a few hundred GeV and several TeV. For a review please see Calibbi et. al {\cite{Calibbi:2006nq,Calibbi:2017uvl}}. For the kind of the operators we are interested in this work, the relevant leptonic number 
 violating process would be $\mu ~\rightarrow~ e$ conversion in Nuclei. This process has been well studied and the limits are reasonably well known (for a review, please see \cite{Kuno:1999jp}), though uncertainties concerning form factors in  various Nuclei persist. The best limit on $\mu \rightarrow e$ comes from the \textbf{SINDRUM II} collaboration, which puts the limits on the Branching fraction to be 
 $\textbf{BR}(\mu \rightarrow e)$ in Ti (Au) is $6.1 (7) \times 10^{-13}$ \cite{SINDRUMII:1998mwd,SINDRUMII:2006dvw}. 

In the following we would like to compare the limits from  $\pi^0 \to \mu^+ e^-$ 
to those from $\mu \to e $ conversion in Nuclei. We do the comparision in the set of low energy operators determined by the LEFT. In this approach, the cLFV would appear as a contact interaction, four Fermi operators at the scale far below the scale of new physics. The operators can be written at the scale of the experiments which is typically far below the scale of the Standard Model. The full effective theory at this scale is called the low energy effective theory (LEFT). This theory is obtained from the Standard Model lagrangian after integrating out the top, Higgs, $W$ and $Z$ particles. The resultant theory contains several operators which could be classified as follows: 
\begin{table}[h!]
    \centering
    \begin{tabular}{|c|c|}
    \hline
    Scalar & Vector\\
    \hline
      $ \mathcal{O}_{eu}^{S RR} =  \dis (\bar{e}_{L i} e_{R j})(\bar{u}_{L w}u_{R t})$  & $\mathcal{O}_{eu}^{V \ LL}  =  \dis (\bar{e}_{L i} \gamma_\mu e_{L j})(\bar{u}_{L w} \gamma^\mu u_{L t})\quad\quad  \mathcal{O}_{eu}^{V \ LR}  =  \dis (\bar{e}_{L i} \gamma_\mu e_{L j})(\bar{u}_{R w} \gamma^\mu u_{R t})$   \\
        $ \mathcal{O}_{eu}^{S RL}  =  \dis (\bar{e}_{L i} e_{R j})(\bar{u}_{R w}u_{L t}) $     & $\mathcal{O}_{eu}^{V \ RR}  =  \dis (\bar{e}_{R i} \gamma_\mu e_{R j})(\bar{u}_{R w} \gamma^\mu u_{R t})\quad\quad \mathcal{O}_{eu}^{V \ RL}  =  \dis (\bar{e}_{R i} \gamma_\mu e_{R j})(\bar{u}_{L w} \gamma^\mu u_{L t})$ \\
        $ \mathcal{O}_{ed}^{S RR} =  \dis (\bar{e}_{L i} e_{R j})(\bar{d}_{L w}d_{R t})$  & $\mathcal{O}_{ed}^{V \ LL}  =  \dis (\bar{e}_{L i} \gamma_\mu e_{L j})(\bar{d}_{L w} \gamma^\mu d_{L t})\quad\quad  \mathcal{O}_{ed}^{V \ LR}  =  \dis (\bar{e}_{L i} \gamma_\mu e_{L j})(\bar{d}_{R w} \gamma^\mu d_{R t})$   \\
        $ \mathcal{O}_{ed}^{S RL}  =  \dis (\bar{e}_{L i} e_{R j})(\bar{d}_{R w}d_{L t}) $     & $\mathcal{O}_{ed}^{V \ RR}  =  \dis (\bar{e}_{R i} \gamma_\mu e_{R j})(\bar{d}_{R w} \gamma^\mu d_{R t})\quad\quad \mathcal{O}_{ed}^{V \ RL}  =  \dis (\bar{e}_{R i} \gamma_\mu e_{R j})(\bar{d}_{L w} \gamma^\mu d_{L t})$ \\
        \hline
    \end{tabular}
    \caption{LEFT semileptonic scalar and vector operators. Notation: $e,u \; \& \; d$ are used for lepton sector, up quark sector and down quark sector respectively. And $i,j,w,t$ are generation indices. For $\pi^0\rightarrow \mu^+ \,\, e^-$ value of i=$\mu$, j=e ,($w,\,\,t$)=(u,u),(d,d)}
    \label{tab:ops}
\end{table}

At the scale of LEFT, these operators can be considered independent. Of course, they could be matched with SMEFT at some high scale and there could be correlations within the operators. In the present work, we assume these operators are independent and derive the limits on them. We find that for the scalar operators, the limits obtained from $\pi^0 \to \mu  e $ can be comparable with those obtained from 
$\mu \to e$ conversion in Ti. For the vector operators, however, the limits from $\mu \to e$ conversion dominate significantly. 
The ratio of the matrix element with the scalar operators to the vector operators is roughly enhanced by a
factor $B_0$ which is a low energy constant appearing in the scalar densities ( in the scalar condensates )
of the chiral perturbation theory. The constant is measured on the lattice and it is typically large $\sim GeV$~\cite{Aoki:2021kgd}. 

The rest of the paper is organised as follows. In the next section, we recap the low energy effective theory and the chiral perturbation theory. In section \ref{sec:LEFT}, we do the matching of the chiral perturbation theory to that of the low energy effective theory at tree level and derive the limits from $BR(\pi^0 \to \mu^+\,e^-) $ in section \ref{sec:matching}. We also derive the limits from $\mu \to e$ conversion. In section \ref{sec:discussion}, we discuss the Leptoquark completion at high scale through a SMEFT matching with LEFT. Thus constraining the parameter space of the Leptoquark models we have considered. We then close with a summary in the last section.

\section{Low-Energy Effective Field Theory}
\label{sec:LEFT}
Low Energy Effective Field Theory (LEFT) \cite{Jenkins:2017dyc,Jenkins:2017jig} is modeled to lie between the Standard Model Theory (SM) scale and the QCD non-perturbative scale. The top quark, the Higgs and the massive gauge bosons are integrated out, and hence LEFT has light quarks, photon and gluon as the dynamical degrees of freedom. Within this framework, running and mixing of effective operators can be computed using perturbative renormalization-group (RG) equations up to the $\Lambda_{QCD}$ scale below which QCD becomes non-perturbative. At which scale the theory must be matched with the $\chi$PT.
There are a total of 80 dimension-6 operators out of which 2 are pure gauge ($X^3$) and 78 are four-fermion contact operators ($\psi^4$). And there exists 6 dimension-5 operators which have mixed fermion gauge structure ($\psi^2 X$). These operators are further divided by their chirality. In general they could be vector, scalar, and tensor like $\psi^4$ operators and are denoted by $V,S,$ and $T$ superscripts.

\subsection{$\chi$PT Lagrangian}
\label{sec:chiPT}
In this subsection we give a brief outline of the chiral Lagrangian. For a complete version we refer the reader to \cite{Gasser:1983yg, Gasser:1984gg}.
In the limit of vanishing quark masses the QCD Hamiltonian has an enhanced symmetry under the chiral group $SU(N_f) \times SU(N_f)$, where $N_f$ is the number of quark flavour. This symmetry is then assumed to be broken by the ground state of the theory to give rise to $N_f^2-1$ Goldstone fields. $\chi$PT is modeled through the addition of quark bilinears using spurion fields which are $N_f \times N_f$ hermitian matrices in flavour space, $v_\mu(x), a_\mu(x), s(x), p(x), t_{\mu\nu}(x)$ associated with vector, axial vector, scalar, pseudo-scalar and tensor currents respectively. The  generating functional for the Greens function corresponding to the vacuum-to-vacuum transition amplitude is then obtained from the Lagrangian,
\bea
\label{eq:chirallagrangian}
\mathcal{L} & =& \dis \mathcal{L}_{\text{QCD}}^{0} + \bar{q}\gamma^\mu \Big(v_\mu(x) + \gamma^5 a_\mu(x)\Big)q - \bar{q} \Big(s(x) - i \gamma^5 p(x)\Big)q \\ \nonumber
&+& \dis \bar{q} \sigma^{\mu\nu} t_{\mu \nu}(x)q \\ \nonumber
\mathcal{L}_{\text{QCD}}^{0} &=& \dis -\frac{1}{2 g_s^2} G_{\mu \nu} G^{\mu \nu} + \bar{q}i \gamma^\mu \Big(\partial_\mu - i G_\mu \Big)q \ ,
\eea
where $G_{\mu \nu}$ is the field strength tensor of the gluon ($G_\mu$). 
For future simplicity, it will be simpler to match with the LEFT operators if we write the above Lagrangian in the chiral notation,
\bea
\label{eq:lagrangian}
\mathcal{L} & =& \dis \mathcal{L}_{\text{QCD}}^{0} + \bar{q}_L\gamma^\mu l_\mu(x) q_L + \bar{q}_R\gamma^\mu r_\mu(x) q_R \\ \nonumber
&+& \dis \bar{q}_L S(x) q_R +  \bar{q}_R S^\dagger(x) q_L \\ \nonumber
&+& \dis \bar{q}_L \sigma^{\mu\nu} t_{\mu \nu}(x)q_R + \bar{q}_L \sigma^{\mu\nu} t_{\mu \nu}^\dagger(x)q_R \ , 
\eea
where 
\bea
\label{eq:chiralnotation}
r_\mu (x) &=& \dis v_\mu(x) + a_\mu(x) \ ,\\ \nonumber
l_\mu (x) &=& \dis v_\mu(x) - a_\mu(x) \ ,\\ \nonumber
S (x) &=& \dis s(x) - i p(x) \ . 
\eea
The three light quark mass matrix are contained in $S$ where are the heavy quarks are integrated out and will not contribute to the low energy effective chiral Lagrangian. 

Considering three flavour, $N_f =3$, the 8 Goldstone fields corresponding to the spontaneous symmetry breaking of $SU(3) \times SU(3)$ to $SU(3)$ is collected in the unitary $3\times3$ matrix $U(x)$. This matrix transforms as 
\[
U'(x) = V_R U(x) V^{\dagger}_L(x),
\]
under the chiral $SU(3) \times SU(3)$. In accordance with the above transformation relation we can write,
$ U(x)= exp\Big(i \frac{\lambda^a \phi^a(x)}{F_0} \Big)$, where $\phi^a$ are the 8 Goldstone fields, $\lambda^a$ the Gell-Mann matrices and $F_0$ the pion decay constant. In the non-linear representation in chiral symmetry $U(x)$ can be written as \cite{Alves:2017avw}
\be
\label{eq:pionmatrix}
U(x) \equiv \exp\Bigg[i \frac{\sqrt{2}}{F_0}\begin{pmatrix}
\frac{\pi^0}{\sqrt{2}}+\frac{\eta}{\sqrt{6}} & \pi^+ & K^+\\
\pi^- & -\frac{\pi^0}{\sqrt{2}}+\frac{\eta}{\sqrt{6}} & K^0\\ 
K^- & \bar{K}^0 & -\frac{2\eta}{\sqrt{6}}
\end{pmatrix}\Bigg]
\ee
The Chiral Lagrangian could be written in terms of these fields as,
\bea
\label{eq:Ulagrangian}
\mathcal{L} &=& \dis \mathcal{L}_1+ \mathcal{L}_2 \\ \nonumber
\mathcal{L}_1 &=& \dis \frac{1}{4}F_0^2\{ tr\Big(D_\mu U^\dagger D^\mu U\Big) + tr\Big(\chi^\dagger U + \chi U^\dagger \Big)\} \ ,\\ \nonumber
\mathcal{L}_2 &=& \dis L_1\langle D^\mu U^\dagger D_\mu U \rangle^2 + L_2 \langle D_\mu U^\dagger D_\nu U \rangle \langle D^\mu U^\dagger D^\nu U \rangle \\ \nonumber
&+& L_3 \langle D^\mu U^\dagger D_\mu U \Big(\chi^\dagger U + U^\dagger \chi\Big)\rangle \\ \nonumber
&+& L_6\langle \chi^\dagger U + \chi U^\dagger \rangle^2 + L_7 \langle \chi^\dagger U - \chi U^\dagger \rangle \\ \nonumber
&+& L_8 \langle \chi^\dagger U \chi^\dagger U + \chi U^\dagger \chi U^\dagger \rangle \\ \nonumber
&-& i L_9 \langle F_{R \mu\nu} D^\mu U D^\nu U^\dagger + F_{L\mu\nu} D^\mu U^\dagger D^\nu U \rangle \\ \nonumber
&+& L_{10} \langle U^\dagger F_{R \mu\nu} U F_L^{\mu\nu} \rangle +H_1 \langle F_{R \mu \nu} F_R^{ \mu \nu} + F_{L\mu \nu} F_L^{ \mu \nu} \rangle + H_2 \langle \chi^\dagger \chi \rangle , 
\eea
where $\langle \, \rangle$ represents the trace of the matrix within and $F^{\mu\nu}_{R} = \partial_\mu r_\nu - \partial_\nu r_\mu - i [ r_\mu, r_\nu]$ and $F^{\mu\nu}_{L} = \partial_\mu l_\nu - \partial_\nu l_\mu - i [ l_\mu, l_\nu]$ are the field strength tensors with $r_\mu$ and $l_\mu $ as defined in Eq.(\ref{eq:chiralnotation}).

Finally, the low energy constants ($L_i$) are fixed from experimental information on D-wave $\pi \pi$ scattering lengths, electro-magnetic charge radius of the pion, decay of $\pi \rightarrow e \nu \gamma$, $K \pi$ scattering and large-$N_c$ arguments (Zweig rule). These are presented in Table \ref{tab:lowenergyconstants}. The contact term $H_2$ has no physical significance, but are needed as counter-terms for renormalization. A more complete expression for a general Lagrangian up to order $p^4$ along with the values computed for the low-energy constants are given in \cite{Gasser:1984gg}.

 \begin{table}[h!]
    \centering
    \begin{tabular}{||c|c||c|c||}
         \hline
        $L_1$ & $(0.9 \pm 0.3 )\times 10^{-3} \times m_{\pi}^+$ & $L_2$  & $(1.7 \pm 0.7 )\times 10^{-3}\times m_{\pi}^+$ \\
        $L_3$ & $(-4.4 \pm 2.5 )\times 10^{-3}\times m_{\pi}^+$ & $L_4$  & $(0 \pm 0.5 )\times 10^{-3}\times m_{\pi}^+$ \\
        $L_5$ & $(2.2 \pm 0.5 )\times 10^{-3}\times m_{\pi}^+$ & $L_6$  & $(0 \pm 0.3 )\times 10^{-3}\times m_{\pi}^+$ \\
        $L_7$ & $(-0.4 \pm 0.15 )\times 10^{-3}\times m_{\pi}^+$ & $L_8$  & $(1.1 \pm 0.3 )\times 10^{-3}\times m_{\pi}^+$ \\
        $L_9$ & $(7.4 \pm 0.7 )\times 10^{-3}\times m_{\pi}^+$ & $L_{10}$  & $(-6.0 \pm 0.7 )\times 10^{-3}\times m_{\pi}^+$ \\
        \hline
    \end{tabular}
    \caption{All values for the low energy coupling constants are taken from \cite{Gasser:1984gg}.}
    \label{tab:lowenergyconstants}
\end{table}

\section{Matching and Lepton flavour violating Pion decay}
\label{sec:matching}

For momentum transfers below $\Lambda_{QCD}$, the quark bilinears in scalar and vector LEFT operators should be matched to $\chi$PT Lagrangian given in Eq.(\ref{eq:Ulagrangian}). Since we are interested in the Lepton flavour violating decays of neutral pion, lets consider the Hamiltonian matrix element
\be
\langle \ell_i\ell_j |\mathcal{O}|\pi^0\rangle \ , 
\label{eq:element}
\ee
where the matrix element $\mathcal{O}$ corresponds to the LEFT operators that contributes to the process. In principle, for a UV complete model, all the operators should contribute with the appropriate UV scale, but for simplicity lets consider the scalar and vectors independently. 

Before proceeding further, a note is in order. 
Not all of the operators contribute to the decay. At tree level only 6 $\psi^4$ type of operators will generate the Lepton flavour violating neutral pion decay. These operators are given in Table \ref{tab:ops}. 
 Regarding tensor operators, for quark bilinears that transforms as a tensor the $\chi$PT replacement rule is given as,
\be
\bar{q}_{Lw}\sigma^{\mu \nu} q_{Rt} = F_1 \langle U_{wt} F_{L}^{\mu \nu} + F_{R}^{\mu \nu} U_{wt}  \rangle + i F_2 \langle D^\mu U_{wr} U^{\dagger}_{rs}D^\nu U_{st} \rangle + \mathcal{O}(p^6) \ .
\label{eq:tensorop}
\ee
On matching with $\chi$PT it can be seen that tensor operators will always have a photon in the final state. and hence will contribute in the process $\pi^0 \rightarrow e^+ e^- \gamma$, for which the world average for the decay fraction,  $\frac{\Gamma(\pi^0 \rightarrow e^+ e^- \gamma)}{\Gamma(\pi^0 \rightarrow \gamma \gamma)} = 1.188 \% $\cite{refId0}, is measured and matched with the theoretical value to $-0.3 \sigma$ precision. 
And hence, tensor operators will not be further discussed here.

\subsection{Scalar operator}

From Table \ref{tab:ops}, there are two scalar operators that contribute to the lepton flavour violating pion decay process. Using these operators, the Hamiltonian given in Eq.(\ref{eq:element}) could be written as, 
\be
\langle \mu e | \Big(\lwc{eu_i}{RL}[S][\mu e  u_i u_i] \mathcal{O}_{eu_i}^{S RL} +\lwc{eu_i}{RR}[S][\mu e  u_i u_i] \mathcal{O}_{eu_i}^{S RR}  \Big)|\pi^0\rangle
\ee
where $\lwc{eu_i}{RL}[S][\mu e  u_i u_i], \; \& \; \lwc{eu_i}{RR}[S][\mu e  u_i u_i] $ are the Wilson Coefficients.  Here, for brevity we have used the notation $u_i = u, \; d$.

At low energies the $\bar{u_i}u_i$ condenses and at order $p^4$ in chiral counting, the quark bilinear in the scalar operators are matched with the chiral Lagrangian. The replacement rule as given in \cite{Dekens:2018pbu,Gasser:1983yg} is,
\bea
\bar{u}_{iL w}u_{iR t} & = &  -2 B_0 \left\{ \frac{1}{4}F_0^2 U_{tw} + L_4 \langle D_\mu U^\dagger D^\mu U \rangle  U_{tw} \right. \nonumber\\
&+& \left. \dis  L_5 (U D_\mu U^\dagger D^\mu U)_{tw} + 2 L_6 \langle U^\dagger \chi + \chi^\dagger U \rangle U_{tw} - 2 L_7  \langle U^\dagger \chi - \chi^\dagger U \rangle U_{tw} \right. \nonumber\\
&+& \left. \dis 2 L_8 ( U \,\chi^\dagger\, U )_{tw} + H_{2}  \,\chi_{tw} \right\} +\mathcal{O}(p^6) \ ,
\label{eq:Chmatching}
\eea
where $F_0$ is the pion decay constant in the chiral limit, $B_0, L_i$ and $H_i$ are the non-perturbative low energy contants and $u_i=u,d$ in our case. $\chi=-2B_0S^{\dagger}$ where `S' is mass matrix $M_{u_i} = diag(m_u,m_d,m_s)$. At leading order $F_0$ and $B_0$ is sufficient to determine the low-energy behaviour. 

 To constraint the Wilson Coefficients responsible for the pion decay process, the terms $\mathcal{O}(p^4)$ in chiral counting are subdominant \cite{Gasser:1984gg}. Hence, keeping the expansion to order $p^2$ in Eq.(\ref{eq:Chmatching}), we get 
\bea
\mathcal{O}_{eu_i}^{S RR} & = & \dis (\bar{\mu}_{L} e_{R})(\bar{u}_{iL w}u_{iR t}) \nonumber\\
& = & \dis (\bar{\mu}_{L} e_{R}) \Big[-2 B_0 \frac{1}{4}F_0^2 U_{tw}\Big] + \mathcal{O}(p^4),
\label{eq:SRR}
\eea
\bea
\mathcal{O}_{eu_i}^{S RL} & = & \dis (\bar{\mu}_{L} e_{R})(\bar{u}_{iR w}u_{iL t}) \nonumber\\
& = & \dis (\bar{\mu}_{L} e_{R}) \Big[-2 B_0 \frac{1}{4}F_0^2 U^{\dagger}_{tw}\Big]+ \mathcal{O}(p^4), 
\label{eq:SRL}
\eea
in $\chi$PT.

To study the effect of operators Eq.(\ref{eq:SRR}) and Eq.(\ref{eq:SRL}) in $\pi^0 \rightarrow \mu e$, we need to expand the matrix in Eq.(\ref{eq:pionmatrix}) to the leading order, $1+i \frac{\lambda^a \phi^a}{F_0}$, and the corresponding Hamiltonian element then becomes,
\bea
\langle e \mu | \Big(\lwc{eu_i}{RL}[S][\mu e  u_i u_i] \mathcal{O}_{eu_i}^{S RL} +\lwc{eu_i}{RR}[S][\mu e  u_i u_i] \mathcal{O}_{eu_i}^{S RR}  \Big)|\pi^0\rangle & = & -\Big(\frac{i B_0F_0}{2}\Big)(\lwc{eu}{RR}[S][\mu e  u u]-\lwc{ed}{RR}[S][\mu e  d d]-\lwc{eu}{RL}[S][\mu e  u u] + \lwc{e d}{RL}[S][\mu e  d d])\nonumber\\
&\times& \dis \langle e \mu |\dis (\bar{\mu}_{L} e_{R})\pi^0|\pi^0\rangle
\label{eq:scalar_element}
\eea

The amplitude, $\mathcal{M}_{\pi^0 \mu^+ e^-}$, for the process computed from the matrix element above, is given as
\bea
\mathcal{M}(\pi^0(p^{\pi^0}) \rightarrow \mu^+(p_{\mu}) e^-(p_e)) & = & \dis -\Big(\frac{i B_0F_0}{2}\Big)(\lwc{eu}{RR}[S][\mu e  u u]-\lwc{ed}{RR}[S][\mu e  d d]-\lwc{eu}{RL}[S][\mu e  u u] + \lwc{e d}{RL}[S][\mu e  d d])\nonumber\\
&\times& \dis \bar{u}(p_e)\frac{1}{2}(1-\gamma^5) v(p_{\mu}) \ .
\label{eq:scalarmatrixelement}
\eea

For the rate, the square of the amplitude is given by,
\bea
\sum_{spin}|\mathcal{M}_{\pi^0 \mu^+ e^-}|^2 & = & \Big|-\Big(\frac{i B_0F_0}{2}\Big)\Big(\lwc{eu}{RR}[S][\mu e  u u]-\lwc{ed}{RR}[S][\mu e  d d]-\lwc{eu}{RL}[S][\mu e  u u] + \lwc{ed}{RL}[S][\mu e  d d]\Big) \Big|^2\nonumber\\
&& \dis \sum_{spin}\Big(\bar{u}(p_e)\frac{1}{2}(1-\gamma^5) v(p_{\mu})\Big)\Big(\bar{u}(p_e)\frac{1}{2}(1-\gamma^5) v(p_{\mu})\Big)^{\dagger}\nonumber\\
&=& \dis \frac{B_0^2}{4}F_0^2 \Big|\Big(\lwc{eu}{RR}[S][\mu e  u u]-\lwc{ed}{RR}[S][\mu e  d d]-\lwc{eu}{RL}[S][\mu e  u u] + \lwc{ed}{RL}[S][\mu e  d d] \Big) \Big|^2 \Big(m_{\pi}^2-(m_e+m_{\mu})^2\Big)
\eea

\begin{table}[]
    \centering
    \begin{tabular}{||c|c||c|c||}
    \hline
         & In GeV  & &In GeV\\
         \hline
        $m_{\pi}^0$ & 134.97$\times 10^{-3}$ & $F_0$  & 92.3$\times 10^{-3}$\cite{Dekens:2018pbu} \\
        $m_{\mu}$  & 105.65$\times 10^{-3}$ & $m_e$ & 0.51$\times 10^{-3}$\\
        $m_u$ & 2.16$\times 10^{-3}$ & $m_d$ & 4.67 $\times 10^{-3}$\\
        $\Gamma(\pi^0\rightarrow \mu^+ e^-)$ & 2.97$\times 10^{-18}$ & $B_0=\frac{m_{\pi^0}^2}{m_u+m_d}$ & 2.667\\
        \hline
    \end{tabular}
    \caption{Input values in GeV. $m_u, m_d$ are MS bar masses at 2 GeV. All values are taken from PDG\cite{Zyla:2020zbs}. }
    \label{tab:input}
\end{table}
Using the masses given in Table \ref{tab:input} and Eq.(\ref{eq:scalarmatrixelement}), the decay width of $\pi^0\rightarrow \mu^+ e^-$ could be directly computed to be,
\begin{eqnarray}
 \Gamma (\pi^0 \rightarrow \mu^+ e^-) &=& \dis \frac{|p_\mu|}{8 \pi m_\pi^2} |\mathcal{M}_{\pi^0 \mu^+ e^-}|^2 \nonumber \\
 &=& \Big|\Big(\lwc{eu}{RR}[S][\mu e  u u]-\lwc{ed}{RR}[S][\mu e  d d]-\lwc{eu}{RL}[S][\mu e  u u] + \lwc{e d}{RL}[S][\mu e  d d]\Big) \Big|^2 (6.007\times 10^{-6})\,\mbox{GeV}^5 \\ \nonumber
& \leq & 2.97 \times 10^{-18}
\end{eqnarray}
And the constraint on Wilson Coefficients becomes,
\be
\Big(\lwc{eu_i}{RR}[S][\mu e  u_i u_i]-\lwc{eu_i}{RL}[S][\mu e  u_i u_i] \Big)\leq
7.03\times 10^{-7}\quad \mbox{GeV}^{-2}
\ee
where the definitions $\lwc{eu_i}{RR}[S][\mu e  u_i u_i]=\lwc{eu}{RR}[S][\mu e  u u]-\lwc{ed}{RR}[S][\mu e  d d]$ and $\lwc{eu_i}{RL}[S][\mu e  u_i u_i]=\lwc{eu}{RL}[S][\mu e  u u] - \lwc{e d}{RL}[S][\mu e  d d]$ have been used.\\ 

The bounds derived so far are at the scale close to mass of the pion and are thus close to QCD scale $\Lambda_{QCD} \sim 2 ~\text{GeV}$. If one would like to see the value of the coupling close to the electro-weak scale, then, one needs to utilise the RGE of the low energy effective theory. These RGE take into consideration the running effects on the couplings between the scales $\sim 2$ GeV to $100$ GeV. The relevant anomalous dimensions are presented in \cite{Jenkins:2017dyc} and reproduced in the Appendix \ref{sec:RGE}. It is understood that these couplings mix under the anomalous dimension matrix, as an example the scalar couplings $\lwc{ed}{RR}[S][1211]$ are mixed with $\lwc{ed}{RR}[S][1222]$ and $\lwc{ed}{RR}[S][1233]$. If all the three couplings $\lwc{ed}{RR}[S][1211], \lwc{ed}{RR}[S][1222], \lwc{ed}{RR}[S][1233]$ are of similar order, then the considered bounds from $\pi \rightarrow \mu e$ on $\lwc{ed}{RR}[S][1211]$ are not significantly affected by the running of the RGE to the high scale. Similar argument also holds for mixing between tensor operators though we have not explicitly given examples of them here. In summary RGE effects do not play an important role and can be safely neglected. 

\begin{figure}[h!]
\centering
\includegraphics[width=0.48\textwidth]{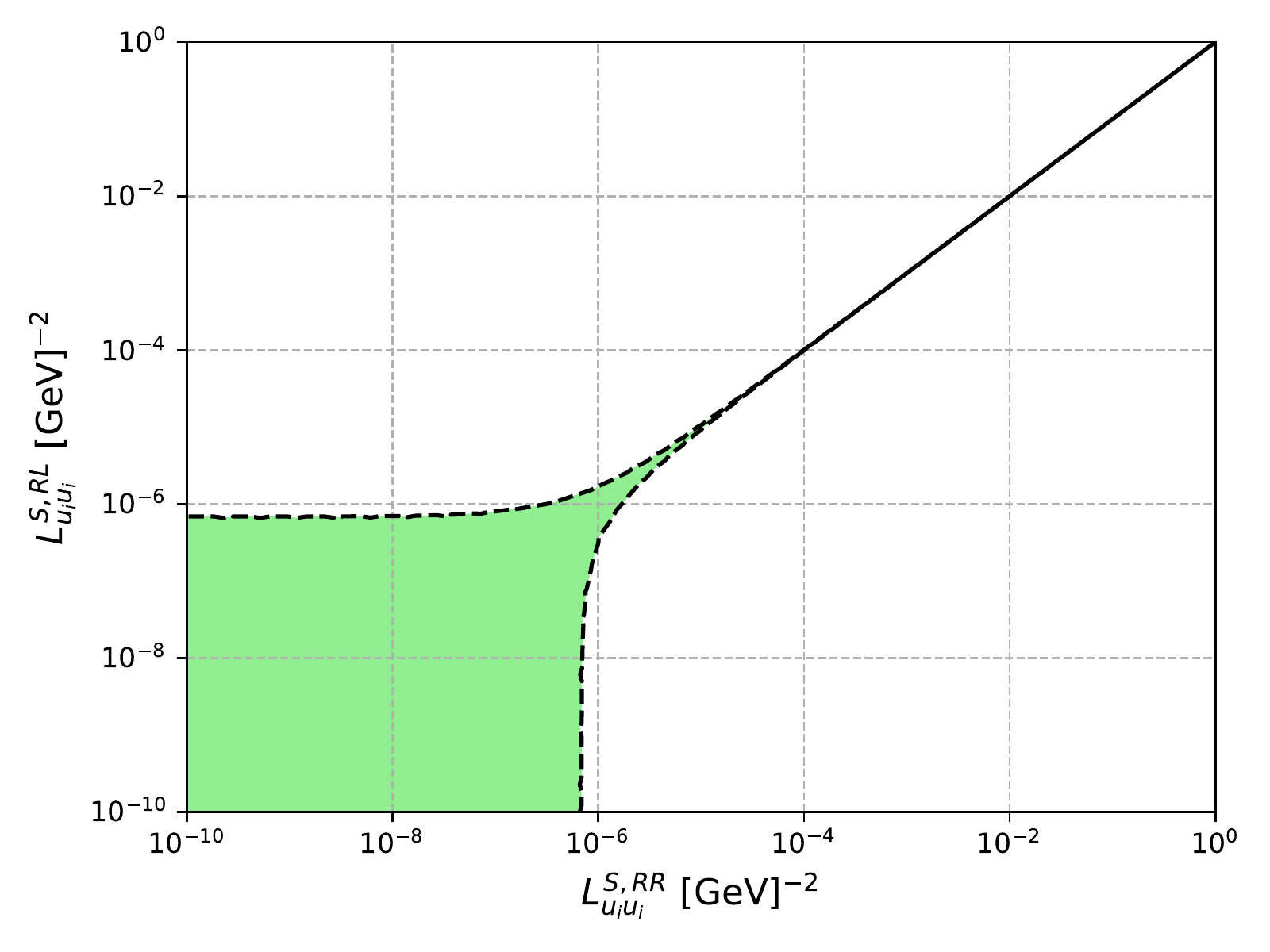}
\includegraphics[width=0.48\textwidth]{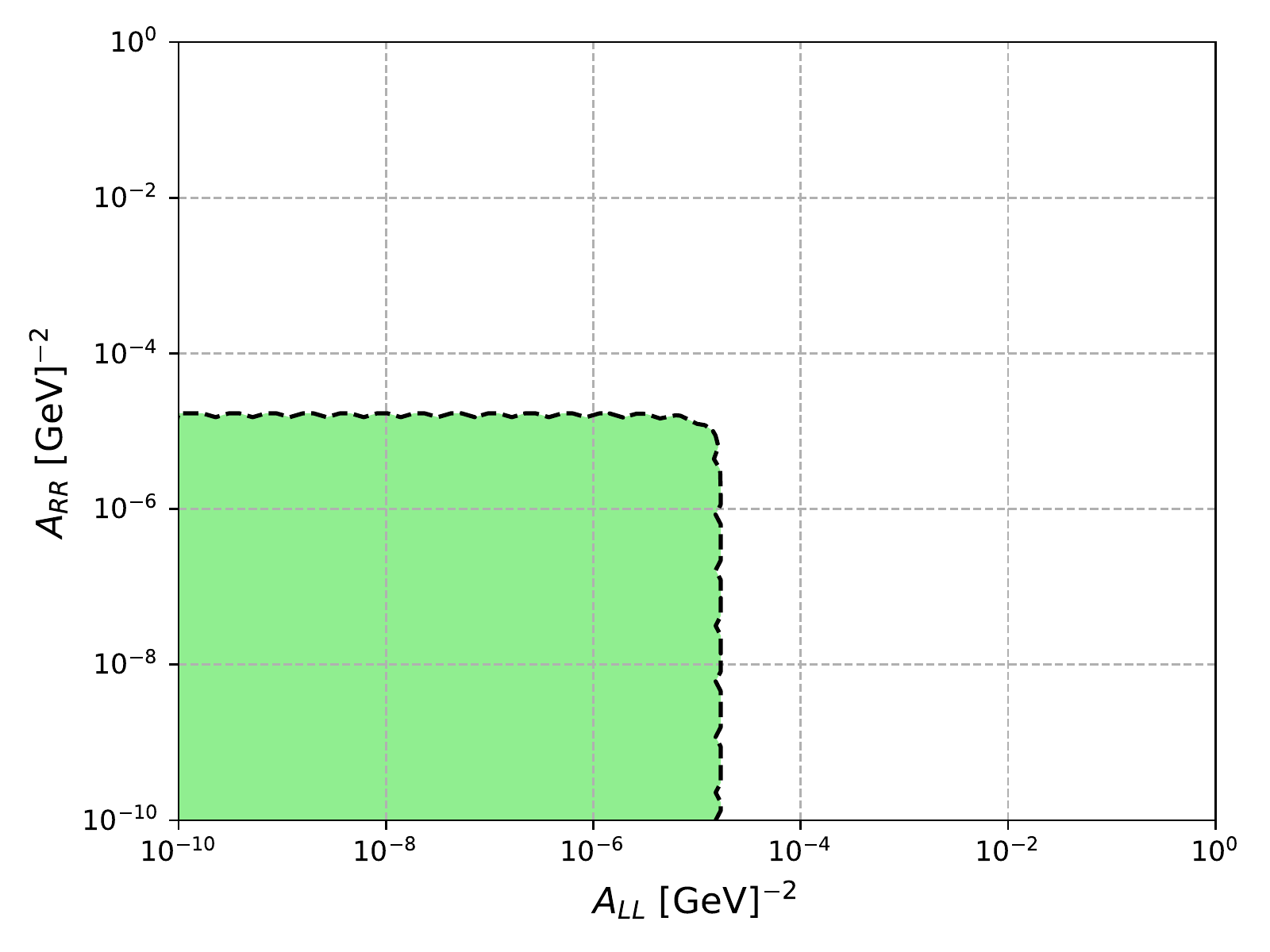}
\caption{On the left hand side, we plot $L^{SRL}_{\mu e u_i u_i} $ vs $ L^{SRR}_{\mu e u_i u_i}$ with the cut-off scale $\Lambda=100$ GeV. In the shaded regions experimental limits from BR($\pi^0 \to \mu e$) are satisfied. On the right hand side, the couplings $A_{RR}$ vs $A_{LL}$ are plotted with the shaded region showing the limits from BR($\pi^0 \to \mu e$).}
\label{fig:opsplot}
\end{figure}

The result could be seen in Fig. \ref{fig:opsplot},
where we have plotted on the left hand side the couplings
$L^{SRL}_{\mu e u_i u_i} $ vs $ L^{SRR}_{\mu e u_i u_i}$. The cut-off scale has been chosen to be 100 GeV. RGE effects have been taken in to consideration while doing this plot. As can be seen, the limits on these couplings 
are $\sim  \text{few} \times 10^{-7}$ (boundary of the shaded region) and weakens  when there are cancellations, along the diagonal of the plot. 

We now proceed to compare the limits from $\mu \to e$ conversion in Nuclei. At the leading order the Br($\mu N \to eN$) is given for scalar operators 
as (in the notation of \cite{Kitano:2002mt}): 
\begin{equation}
    Br(\mu N \to e N) = \frac{2 G_F^2 (G_S^{(d,p)} S^{(p)} + G_S^{(d,n)} S^{(n)})^2 (|g_{LS(d)}|^2
    + |g_{RS(d)}|^2)}  {w_{capt}}, 
\end{equation}
where $G_F$ stands for the Fermi constant and $G_S^{(d,p)}, S^{(p)},  G_S^{(d,n)}, S^{(n)}$ correspond to various form factors and number densities represented in terms of overlap integrals. $w_{capt}$ is the capture rate.  Using the values quoted for $Ti$
we have the limit on the scalar couplings to be:
\begin{equation}
   \sqrt{ (|g_{LS(d)}|^2 +  |g_{RS(d)}|^2)} \leq 7.8 \times 10^{-7}.
\end{equation}
The limit is of similar order of magnitude as one gets from Pion LFV decay.

\subsection{Vector operators}
The NP can also contribute to the process $\pi^0 \rightarrow \mu^+ e^-$ through effective vector operators.The corresponding Lepton Flavour violating Hamiltonian element is,
\be
\langle \ell_i\ell_j | \mathcal{O}|\pi^0\rangle \ ,
\ee
where $\mathcal{O}$ is now the vector operators given in Table \ref{tab:ops}. Using these operators the above matrix element could be written as,

\be
\langle e \mu | \Big(\lwc{eu_i}{LL}[V][\mu e  u_i u_i] \mathcal{O}_{eu_i}^{V LL} +\lwc{eu_i}{LR}[V][\mu e  u_i u_i] \mathcal{O}_{eu_i}^{V LR} +\lwc{eu_i}{RR}[V][\mu e  u_i u_i] \mathcal{O}_{eu_i}^{V RR} +\lwc{eu_i}{RL}[V][\mu e  u_i u_i] \mathcal{O}_{eu_i}^{V RL}  \Big)| \pi^0\rangle
\ee
where $\lwc{eu_i}{LL}[V][\mu e  u_i u_i] , \lwc{eu_i}{LR}[V][\mu e  u_i u_i] $, $\lwc{eu_i}{RR}[V][\mu e  u_i u_i] $ and $\lwc{eu_i}{RL}[V][\mu e  u_i u_i] $ are the Wilson Coefficient. And as before $u_i = u, \; d$. 

Like for the scalar operators, at low energies the $\bar{u_i}u_i$ condenses and then quark bilinear in the vector operators should be matched with the chiral Lagrangian. The replacement rule for vector LEFT quark bilinear could be written as, 
\bea
\bar{u_i}_{L} \gamma^{\mu}u_{iL}&=& \frac{i}{2}F_0^2\langle D_\mu U U^\dagger\rangle + 4 i L_1 \vev{D_\nu U^\dagger D^\nu U} \vev{ D_\mu U U^\dagger  } + 4i L_2 \vev{D^\mu  U^\dagger D^\nu U} \vev{ D_\nu U U^\dagger }  \nonumber\\
&+& \dis 2 i L_3 \vev{\left(U^\dagger  D^\mu U - D^\mu U^\dagger    U  \right) D_\nu U^\dagger D^\nu U } + 2 i L_4 \vev{  D_\mu U U^\dagger   } \vev{U^\dagger \chi + \chi^\dagger U}  \nonumber\\
&+& \dis i L_5  \vev{ \left( U^\dagger  D^\mu U  -D^\mu U^\dagger  U  \right) (U^\dagger \chi + \chi^\dagger U)  } \nonumber\\
&+& \dis L_9 \left[ -\vev{ F_R^{\mu \nu} D_\nu U U^\dagger}  - \vev{ U D_\nu U^\dagger F_R^{\mu \nu} } + \vev{ D_\nu U F_L^{\mu \nu} U^\dagger} + \vev{ U F_L^{\mu \nu} D_\nu U^\dagger}  \right]  \nonumber\\
&-& \dis i L_9 \vev{ D^\nu(D_\mu U D_\nu U^\dagger-D_\nu U D_\mu U^\dagger}  + 2 L_{10} \vev{ D_\nu( U F_L^{\mu \nu} U^\dagger)}  \nonumber\\
&+& \dis  4 H_1 \vev{ D_\nu F_R^{\mu \nu}} + \epsilon\text{ terms}+ \mathcal{O}(p^6)  \ ,
\label{vectormatching}
\eea
where $F^{\mu\nu}_{R}$ and $F^{\mu\nu}_{L}$ are the field strength tensors as defined previously.  Similarly $\bar{u}_{iR} \gamma^{\mu}u_{iR}$ matching is same as in Eq.(\ref{vectormatching}) by replacing $U\rightarrow U^{\dagger}$, $\chi\rightarrow \chi^{\dagger}$.

To the most dominant term, contributing to the process $\pi^0 \rightarrow \mu^+ e^-$, in the order in expansion of chiral Lagrangian becomes,
\bea
\mathcal{O}_{eu_i}^{V LL}&=& \dis (\bar{\mu}_{L} \gamma_\mu e_{L})(\bar{u}_{iL w} \gamma^\mu u_{iL t})\quad\quad\quad\quad  \mathcal{O}_{eu_i}^{V LR}  =  \dis (\bar{\mu}_{L} \gamma_\mu e_{L})(\bar{u}_{iR w} \gamma^\mu u_{iR t})\nonumber\\
&=&\dis (\bar{\mu}_{L} \gamma^\mu e_{L})\Big( \frac{i}{2}F_0^2( D_\mu U U^\dagger)_{tw}\Big)\quad\quad\quad\,\,\,=\dis  (\bar{\mu}_{L} \gamma^\mu e_{L})\Big( \frac{i}{2}F_0^2( D_\mu U^\dagger U)_{tw}\Big)
\label{eq:vectoropsLL}
\eea
and
\bea
\mathcal{O}_{eu_i}^{V RL}&=& \dis (\bar{\mu}_{R} \gamma_\mu e_{R})(\bar{u}_{iL w} \gamma^\mu u_{iL t})\quad\quad\quad\quad  \mathcal{O}_{eu_i}^{V LR}  =  \dis (\bar{\mu}_{R} \gamma_\mu e_{R})(\bar{u}_{iR w} \gamma^\mu u_{iR t})\nonumber\\
&=&\dis (\bar{\mu}_{R} \gamma^\mu e_{R})\Big( \frac{i}{2}F_0^2( D_\mu U U^\dagger)_{tw}\Big)\quad\quad\quad\,\,\,=\dis  (\bar{\mu}_{R} \gamma^\mu e_{R})\Big( \frac{i}{2}F_0^2( D_\mu U^\dagger U)_{tw}\Big)
\label{eq:vectoropsRL}
\eea
where the neutral pion arises from $(D_{\mu}UU^{\dagger})_{uu}\rightarrow\frac{i}{F_0}\partial_{\mu}\pi^0, \quad (D_{\mu}UU^{\dagger})_{dd}\rightarrow-\frac{i}{F_0}\partial_{\mu}\pi^0$ and $(D_{\mu}U^{\dagger}U)_{uu}\rightarrow-\frac{i}{F_0}\partial_{\mu}\pi^0, \ \text{and} \  (D_{\mu}U^{\dagger}U)_{dd}\rightarrow\frac{i}{F_0}\partial_{\mu}\pi^0 $, in the expansion of $U(x)$ given in Eq.(\ref{eq:chiralnotation}) up to first order in $\frac{1}{F_0}$. 

Then, to study the effect of the operators in Eq.(\ref{eq:vectoropsLL}) and Eq.(\ref{eq:vectoropsRL}) in the Lepton flavour violating neutral pion decay process, the matrix element could be derived as,
\bea
\langle e \mu | \Big(\lwc{eu_i}{LL}[V][\mu e  u_i u_i] \mathcal{O}_{eu_i}^{V LL} +\lwc{eu_i}{LR}[V][\mu e  u_i u_i] \mathcal{O}_{eu_i}^{V LR} +\lwc{eu_i}{RR}[V][\mu e  u_i u_i] \mathcal{O}_{eu_i}^{V RR} +\lwc{eu_i}{RL}[V][\mu e  u_i u_i] \mathcal{O}_{eu_i}^{V RL}  \Big)| \pi^0\rangle &=& \dis \\ \nonumber
\langle e \mu |\Big(\lwc{eu_i}{LR}[V][\mu e  u_i u_i]-\lwc{eu_i}{LL}[V][\mu e  u_i u_i]\Big)(\bar{\mu}_{L} \gamma^\mu e_{L})\frac{F_0}{2}\partial_{\mu}\pi^0|\pi^0\rangle &+& \dis\\ \nonumber
 \langle e \mu |\Big(\lwc{eu_i}{RR}[V][\mu e  u_i u_i]-\lwc{eu_i}{RL}[V][\mu e  u_i u_i]\Big)(\bar{\mu}_{R} \gamma^\mu e_{R})\frac{F_0}{2}\partial_{\mu}\pi^0|\pi^0\rangle
\eea
where $\lwc{eu_i}{AB}[V][\mu e  u_i u_i]=\lwc{eu}{AB}[V][\mu e  u u]-\lwc{ed}{AB}[V][\mu e  d d]$. 
For simplicity, lets define 
\bea
A_{LL}=\lwc{eu_i}{LR}[V][\mu e  u_i u_i]-\lwc{eu_i}{LL}[V][\mu e  u_i u_i]\\ \nonumber
A_{RR}=\lwc{eu_i}{RR}[V][\mu e  u_i u_i]-\lwc{eu_i}{RL}[V][\mu e  u_i u_i]
\eea

The corresponding amplitude can be written as
\bea
\mathcal{M}(\pi^0(p^{\pi^0}) \rightarrow \mu^+(p_{\mu}) e^-(p_e)) & = & \dis \frac{F_0}{2}\Big(A_{LL}(\bar{u}(p_e) \gamma^\mu\frac{(1-\gamma^5)}{2} v(p_{\mu}))p_{\mu}^{\pi_0}\nonumber\\
&+&A_{RR}(\bar{u}(p_e) \gamma^\mu\frac{(1+\gamma^5)}{2} v(p_{\mu}))p_{\mu}^{\pi_0}\Big) \ .
\label{eq:vectormatrixelement}
\eea

Squaring this amplitude we get,
\bea
\sum_{spin}|\mathcal{M}_{\pi\mu e}|^2 & = & \dis \frac{F_0^2}{4}\Big[2(|A_{LL}|^2+|A_{RR}|^2)\frac{1}{2}\Big(m_{\pi^0}^2(m_\mu^2+m_e^2)-(m_\mu^2-m_e^2)^2\Big)\Big] 
\eea

In limit $m_e\rightarrow 0$, using masses in Table \ref{tab:input} and Eq.(\ref{eq:vectormatrixelement}),
the decay width of neutral pion decay process could be computed to be,
\begin{eqnarray}
\Gamma (\pi^0 \rightarrow \mu^+ e^-) &=& \dis \frac{|\vec{p}_{\mu}|}{8 \pi m_{\pi^0}^2} \sum_{spin}|\mathcal{M}_{\pi^0 \mu^+ e^-}|^2 \nonumber \\
 &=& \dis (|A_{LL}|^2+|A_{RR}|^2)\times (9.57\times 10^{-9})\,\,(\mbox{GeV})^5
\end{eqnarray}
This puts a constraint on the combination of the Wilson Coefficients as
\bea
(|A_{LL}|^2+|A_{RR}|^2)&\leq&  3.104\times 10^{-10} \,\, (\mbox{GeV})^{-4}
\eea

We now compare the limits for the same from $\mu \to e$ conversion in Ti. Following \cite{Kitano:2002mt} we have: 
\begin{equation}
    Br(\mu N \to e N) = \frac{2 G_F^2 (V^{(p)})^2(|\bar{g}_{LV}^{(p)}|^2 +  |\bar{g}_{RV}^{(p)}|^2)}  {w_{capt}}, 
\end{equation}
where $V^{(p)}$ represents vectorial form factors. For Ti, we find the limits to be 
\begin{equation}
(|\bar{g}_{LV}^{(p)}|^2 + |\bar{g}_{RV}^{(p)}|^2) < 4.0 \times 10^{-14},
\end{equation}
which is far stronger than the one from Pion LFV decay. In Fig. \ref{fig:opsplot} we show limit (shaded region) in the plane of $A_{RR}$
and $A_{RL}$, as can be seen the limit
is of the order of $\sim 10^{-4}$ which becomes weaker along the diagonal due to cancellations. As mentioned before RGE corrections are taken in to consideration while plotting this graph. 

\section{Discussion}
\label{sec:discussion}

Though the results are be adaptable for any BSM NP model, in this section we relie on Lepto-Quarks to generate the flavour violating coupling. 
Renormalisable Lagrangian of scalar and vector Lepto-Quarks that couple to SM without generating tree level proton decay operators\cite{Assad:2017iib} could be written as 
\bea
\mathcal{L}_{\Delta} &=& \dis \Big(\lambda^{i j}_{Lu} \bar{L}_L^{i}u^j_R + \lambda^{i j}_{eQ} \bar{e}^i_R i \tau_2 Q^j_L\Big)\Delta_{-\frac{7}{6}} \\
\mathcal{L}_V &=& \dis \Big( g_{LQ} \bar{L}_L \gamma_\mu Q_L + g_{ed} \bar{e}_R \gamma_\mu d_R\Big) V_{-2/3}^\mu + g_{eu} \bar{e}_R \gamma_\mu u_R V_{-5/3}^\mu + g'_{LQ} \bar{L}_L \gamma_\mu \tau^a Q_L V_{-2/3}^{\mu \ a} 
\eea
where $\Delta_{-7/6}$ is the scalar and $V^{\mu}_{-2/3},V^{\mu}_{-5/6}, V^{\mu \ a}_{-2/3},$ are the vector Lepto-Quarks. 

Using the above Lagrangian, let us discuss the tree level matching of the LEFT Wilson coefficients that contribute to the $\pi^0 \rightarrow \mu e$ process with the Standard Model Effective Field Theory (SMEFT) operators generated from the model.  

{\bf Scalar operators} :

At tree level, 
\bea
\lwc{eu}{RR}[S][\mu e  u u] = -C^{(1)}_{lequ}(\mu euu) = \frac{\lambda^{i j}_{Lu} \lambda^{*i j}_{eQ}}{M_{\Delta_{7/6}}^2} \ , \lwc{ed}{RR}[S][\mu e  d d]  = 0
\eea

At tree level, 
\bea
\lwc{ed}{RL}[S][\mu e  dd]  = \sum_{i=d,s,b} V_{di}C_{ledq}(\mu e di) = 0 \ , \lwc{eu}{RL}[S][\mu e  u u]  = 0
\eea

Then, from the constraint equation 
\be
\Big(\lwc{eu_i}{RR}[S][\mu e  u_i u_i]-\lwc{eu_i}{RL}[S][\mu e  u_i u_i]\Big)\leq
7.03\times 10^{-7}\quad \mbox{GeV}^{-2}
\ee
we get
\be
\frac{\lambda^{i j}_{Lu} \lambda^{*i j}_{eQ}}{M_{\Delta_{7/6}}^2} \leq
7.03\times 10^{-7}\quad \mbox{GeV}^{-2}
\ee

{\bf Vector Fields} :

At tree level, 
\bea
\lwc{eu}{LL}[V][\mu e  u u] = C^{(1)}_{lq}(\mu e q_1 q_1)-C^{(3)}_{lq}(\mu e q_1 q_1 ) = 0
\eea

At tree level, assuming Lepto-Quark coupling to be universal, 
\bea
\lwc{ed}{LL}[V][\mu e  dd]  = \sum_{i,j=d,s,b}\Big(C^{(1)}_{lq}(\mu e i j) + C^{(3)}_{lq}(\mu e i j)\Big) V_{jd} V^\dagger_{di} = \frac{g_{LQ}g^*_{ed}}{M_{V_{2/3}}^2}
\eea

At tree level, 
\bea
\lwc{eu}{LR}[V][\mu e  u u]  = C_{lu}(\mu e u u) = 0 \ , C_{\mu e d d }^{V LR} = \lwc{ed}{LL}[V][\mu e  dd] = 0
\eea

At tree level, 
\bea
\lwc{eu}{RR}[V][\mu e  u u]  = C_{eu}(\mu e u u) = \frac{g_{eu}g^*_{eu}}{M_{V_{5/3}}^2} \ , \lwc{ed}{RR}[V][\mu e  d d]  = C_{ed}(\mu e d d) = \frac{g_{ed}g^*_{ed}}{M_{V_{2/3}}^2}
\eea

At tree level, 
\bea
\lwc{eu}{RL}[V][\mu e  u u]  = C_{qe}(\mu e u u)=0 \ , \lwc{ed}{RL}[V][\mu e  d d]  = C_{qe}(\mu e d d)=0
\eea

From the constraint on the vector LEFT operator's Wilson coefficients
\bea
(|A_{LL}|^2+|A_{RR}|^2)&\leq&  3.104\times 10^{-10} \,\, (\mbox{GeV})^{-4} , 
\eea
where
\bea
A_{LL}=\lwc{eu_i}{LR}[V][\mu e  u_i u_i]-\lwc{eu_i}{LL}[V][\mu e  u_i u_i]\\ \nonumber
A_{RR}=\lwc{eu_i}{RR}[V][\mu e  u_i u_i]-\lwc{eu_i}{RL}[V][\mu e  u_i u_i]
\eea
we get,
\be
\Big|\frac{g_{LQ}g^*_{ed}}{M_{V_{2/3}}^2}\Big|^2 + \Big| \frac{g_{eu}g^*_{eu}}{M_{V_{5/3}}^2}- \frac{g_{ed}g^*_{ed}}{M_{V_{2/3}}^2}\Big|^2  \leq 3.104\times 10^{-10} \,\, (\mbox{GeV})^{-4}
\ee
In Table \ref{tab:LQ}, we give the summary of the matching done at $M_W$ scale, for detailed calculation refer to the Appendix \ref{sec:SMEFT}.

\begin{table}[h!]
\centering
\begin{tabular}{|c|c|c|}
\hline
LeptoQuark Model & Coupling  & Limit   \\
\midrule\midrule
Scalar LQ ($\Delta_{-\frac{7}{6}}$)  &  $\frac{\lambda^{i j}_{Lu} \lambda^{*i j}_{eQ}}{M_{\Delta_{7/6}}^2}$&$ \leq
7.03\times 10^{-7}\quad \mbox{GeV}^{-2}$
 \\
Vector LQ ($V^\mu_{-\frac{2}{3}}$, $V^\mu_{-\frac{5}{3}}$)  & $
\Big|\frac{g_{LQ}g^*_{ed}}{M_{V_{2/3}}^2}\Big|^2 + \Big| \frac{g_{eu}g^*_{eu}}{M_{V_{5/3}}^2}- \frac{g_{ed}g^*_{ed}}{M_{V_{2/3}}^2}\Big|^2 $ & $ \leq 3.104\times 10^{-10} \,\, (\mbox{GeV})^{-4}$\\
\hline
\end{tabular}
\label{tab:LQ}
\caption{Limits on first two generation flavour violating couplings of Scalar and Vector LeptoQuarks}
\end{table}


\subsection{Summary}
In this paper we derived the model independent bounds on Lepton flavour violating LEFT operators in neutral pion decay process which involves only the first generation quarks. By using LEFT-$\chi$PT matching it is clear that only scalar and vector operators contribute to the process. Tensor operators, as shown in Eq.(\ref{eq:tensorop}), will always result in a photon emission along with the dileptons. 
We have also shown that constraint on scalar operators are much stringent than the one on vector operators. This is because of the scalar amplitudes are enhanced by the condesation parameter $B_0$ compared to the vector ones. 

 It is generally believed that $\mu \to e$ conversion in Nuclei gives the strong bounds on the first two generational
lepton number/flavour violating operators of the $q^2l^2$ type. While this is generally true, the limits on the process $\pi^0 \to \mu e $ also give comparable constraints on scalar $q^2 l^2$ operators for vector operators, the limits are weaker than those from $\mu \to e$ conversion.

\begin{acknowledgments}
\noindent
 We are thankful to  B. Ananthanarayan and G. D'Ambrosio for extensive discussions. M.T.A. acknowledges financial support of DST through INSPIRE Faculty grant [DST/INSPIRE/04/2019/002507]. SKV acknowledges support from the Institute of Excellence grants from IISc. P.L. partially
supported by the National Science Centre, Poland, under research grant 2017/26/E/ST2/00135.
\end{acknowledgments}
\section{Appendix}
\subsection{RGE}
\label{sec:RGE}
The needed LEFT Renormalization Group Equations \cite{Jenkins:2017dyc} for the operators given in Table \ref{tab:ops} is given as,
\bea
	\dlwc{eu}{RR}[S][prst] &= - \left[ 6 e^2 (\q_e^2 + \q_u^2) + 6 g^2 C_F \right] \lwc{eu}{RR}[S][prst] -96 e^2 \q_e \q_u \lwc{eu}{RR}[T][prst]  -192 e^2 \q_e \q_u \lwc{e \gamma}{}[][pr] \lwc{u \gamma}{}[][st]   \, , \\
\nnn
	\dlwc{ed}{RR}[S][prst] &= - \left[ 6 e^2 (\q_e^2 + \q_d^2) + 6 g^2 C_F \right] \lwc{ed}{RR}[S][prst] -96 e^2 \q_e \q_d \lwc{ed}{RR}[T][prst]  -192 e^2 \q_e \q_d \lwc{e \gamma}{}[][pr] \lwc{d \gamma}{}[][st]   \, , \\
\nnn
    \dlwc{eu}{RL}[S][prst] &= - \left[ 6 e^2 (\q_e^2 + \q_u^2) + 6 g^2 C_F \right] \lwc{eu}{RL}[S][prst]  -192 e^2 \q_e \q_u \lwc{e \gamma}{}[][pr] \lwc{u \gamma}{}[][st]   \, , \\
\nnn
	\dlwc{ed}{RL}[S][prst] &= - \left[ 6 e^2 (\q_e^2 + \q_d^2) + 6 g^2 C_F \right] \lwc{ed}{RL}[S][prst]  -192 e^2 \q_e \q_d \lwc{e \gamma}{}[][pr] \lwc{d \gamma}{}[][st]   \, , \\
\nnn
	\dlwc{eu}{LL}[V][prst] &= \frac{4}{3} e^2 \q_e \delta_{pr} \begin{aligned}[t]
			&\bigg[ N_c \q_d \Bigl(\lwc{ud}{LL}[V][stww]+\lwc{ud}{LR}[V][stww] \Bigr) + N_c \q_u \Bigl( 2 \lwc{uu}{LL}[V][stww]+\lwc{uu}{LR}[V][stww] \Bigr) \nn
			& + \q_e \Bigl( \lwc{eu}{LL}[V][wwst]+\lwc{ue}{LR}[V][stww] \Bigr) + 2 \q_u \lwc{uu}{LL}[V][swwt] \bigg] \end{aligned} \nn
		&\quad + \frac{4}{3} e^2 \q_u \delta_{st} \begin{aligned}[t]
			& \bigg[ N_c \q_d \Bigl( \lwc{ed}{LL}[V][prww]+\lwc{ed}{LR}[V][prww] \Bigr) + N_c \q_u \Bigl( \lwc{eu}{LL}[V][prww]+\lwc{eu}{LR}[V][prww] \Bigr) \nn
			& + \q_e \Bigl( 4 \lwc{ee}{LL}[V][prww]+\lwc{ee}{LR}[V][prww] \Bigr) \bigg] \end{aligned} \nn
		&\quad + 12 e^2 \q_e \q_u \lwc{eu}{LL}[V][prst]  - \frac{16}{3} C_F e^2 \q_e \q_u \lwc{u G}{}[][sw] \lwc{u G}{*}[][tw] \delta_{pr}  + 8 C_F e g \q_e \Bigl( \lwc{uG}{}[][sw] \lwc{u \gamma}{*}[][tw] + \lwc{u \gamma}{}[][sw] \lwc{uG}{*}[][tw] \Bigr) \delta_{pr}  \nn
		&\quad + e^2  \Bigl( 24 \q_u^2 + \frac{32}{3} \q_e \q_u \Bigr)  \lwc{e \gamma}{}[][pw] \lwc{e \gamma}{*}[][rw] \delta_{st} + e^2 \Bigl( 24 \q_e^2 + \frac{32}{3} \q_e \q_u \Bigr)   \lwc{u \gamma}{}[][sw] \lwc{u \gamma}{*}[][tw] \delta_{pr} \nn
		&\quad + 2 e^2 \q_e \q_u   \zeta_e  \delta_{pr} \delta_{st} \, , \\
\nnn
	\dlwc{ed}{LL}[V][prst] &= \frac{4}{3} e^2 \q_e \delta_{pr} \begin{aligned}[t]
			& \bigg[ N_c \q_u \Bigl( \lwc{ud}{LL}[V][wwst]+\lwc{du}{LR}[V][stww] \Bigr) + N_c \q_d \Bigl( 2 \lwc{dd}{LL}[V][stww]+\lwc{dd}{LR}[V][stww] \Bigr) \nn
			& + \q_e \Bigl( \lwc{ed}{LL}[V][wwst]+\lwc{de}{LR}[V][stww] \Bigr) + 2 \q_d \lwc{dd}{LL}[V][swwt] \bigg] \end{aligned} \nn
		&\quad + \frac{4}{3} e^2 \q_d \delta_{st} \begin{aligned}[t]
			& \bigg[ N_c \q_d \Bigl( \lwc{ed}{LL}[V][prww]+\lwc{ed}{LR}[V][prww] \Bigr) + N_c \q_u \Bigl( \lwc{eu}{LL}[V][prww]+\lwc{eu}{LR}[V][prww] \Bigr) \nn
			& + \q_e \Bigl( 4 \lwc{ee}{LL}[V][prww]+\lwc{ee}{LR}[V][prww] \Bigr) \bigg] \end{aligned} \nn
		&\quad + 12 e^2 \q_d \q_e \lwc{ed}{LL}[V][prst]  - \frac{16}{3} C_F e^2 \q_e \q_d \lwc{d G}{}[][sw] \lwc{d G}{*}[][tw] \delta_{pr}  + 8 C_F e g \q_e \Bigl( \lwc{dG}{}[][sw] \lwc{d \gamma}{*}[][tw] + \lwc{d \gamma}{}[][sw] \lwc{dG}{*}[][tw] \Bigr) \delta_{pr}  \nn
		&\quad + e^2  \Bigl( 24 \q_d^2 + \frac{32}{3} \q_e \q_d \Bigr)  \lwc{e \gamma}{}[][pw] \lwc{e \gamma}{*}[][rw] \delta_{st}  +  e^2  \Bigl( 24 \q_e^2 + \frac{32}{3} \q_e \q_d \Bigr)   \lwc{d \gamma}{}[][sw] \lwc{d \gamma}{*}[][tw] \delta_{pr} \nn
		&\quad + 2 e^2 \q_e \q_d   \zeta_e  \delta_{pr} \delta_{st}  \, ,
\nnn
\eea
\bea
	\dlwc{eu}{LR}[V][prst] &= \frac{4}{3} e^2 \q_e \delta_{pr} \begin{aligned}[t]
			& \bigg[ N_c \q_d \Bigl(\lwc{du}{LR}[V1][wwst]+\lwc{ud}{RR}[V1][stww]\Bigr) + N_c \q_u \Bigl(\lwc{uu}{LR}[V1][wwst]+2\lwc{uu}{RR}[V][stww] \Bigr) \nn
			& + \q_e \Bigl(\lwc{eu}{LR}[V][wwst]+\lwc{eu}{RR}[V][wwst]\Bigr) + 2 \q_u \lwc{uu}{RR}[V][swwt] \bigg] \end{aligned} \nn
		&\quad + \frac{4}{3} e^2 \q_u \delta_{st} \begin{aligned}[t]
			& \bigg[ N_c \q_d \Bigl(\lwc{ed}{LL}[V][prww]+\lwc{ed}{LR}[V][prww]\Bigr) + N_c \q_u \Bigl(\lwc{eu}{LL}[V][prww]+\lwc{eu}{LR}[V][prww]\Bigr) \nn
			& + \q_e \Bigl( 4 \lwc{ee}{LL}[V][prww]+\lwc{ee}{LR}[V][prww]\Bigr) \bigg] \end{aligned} \nn
		&\quad - 12 e^2 \q_e \q_u \lwc{eu}{LR}[V][prst]   - \frac{16}{3} C_F e^2 \q_e \q_u \lwc{uG}{*}[][ws] \lwc{uG}{}[][wt] \delta_{pr}  + 8 C_F e g \q_e \Bigl( \lwc{uG}{*}[][ws] \lwc{u\gamma}{}[][wt] + \lwc{u\gamma}{*}[][ws] \lwc{uG}{}[][wt] \Bigr) \delta_{pr}  \nn
		&\quad + e^2 \Bigl( \frac{32}{3} \q_e \q_u - 24 \q_u^2 \Bigr)  \lwc{e \gamma}{}[][pw] \lwc{e \gamma}{*}[][rw] \delta_{st}  + e^2 \Bigl( \frac{32}{3} \q_e \q_u - 24 \q_e^2 \Bigr)  \lwc{u \gamma}{*}[][ws] \lwc{u \gamma}{}[][wt] \delta_{pr}  \nn
		&\quad + 2 e^2 \q_u \q_e \zeta_e   \delta_{pr} \delta_{st}  \, , \\
\nnn
	\dlwc{ed}{LR}[V][prst] &= \frac{4}{3} e^2 \q_e \delta_{pr} \begin{aligned}[t]
			& \bigg[ N_c \q_d \Bigl(\lwc{dd}{LR}[V1][wwst]+2\lwc{dd}{RR}[V][stww]\Bigr) + N_c \q_u \Bigl(\lwc{ud}{LR}[V1][wwst]+\lwc{ud}{RR}[V1][wwst] \Bigr) \nn
			& + \q_e \Bigl(\lwc{ed}{LR}[V][wwst]+\lwc{ed}{RR}[V][wwst]\Bigr) + 2 \q_d \lwc{dd}{RR}[V][swwt] \bigg] \end{aligned} \nn
		&\quad + \frac{4}{3} e^2 \q_d \delta_{st} \begin{aligned}[t]
			& \bigg[ N_c \q_d \Bigl(\lwc{ed}{LL}[V][prww]+\lwc{ed}{LR}[V][prww]\Bigr) + N_c \q_u \Bigl(\lwc{eu}{LL}[V][prww]+\lwc{eu}{LR}[V][prww] \Bigr) \nn
			& + \q_e \Bigl( 4 \lwc{ee}{LL}[V][prww]+\lwc{ee}{LR}[V][prww]\Bigr) \bigg] \end{aligned} \nn
		&\quad - 12 e^2 \q_d \q_e \lwc{ed}{LR}[V][prst]  - \frac{16}{3} C_F e^2 \q_e \q_d \lwc{dG}{*}[][ws] \lwc{dG}{}[][wt] \delta_{pr}  + 8 C_F e g \q_e \Bigl( \lwc{dG}{*}[][ws] \lwc{d\gamma}{}[][wt] + \lwc{d\gamma}{*}[][ws] \lwc{dG}{}[][wt] \Bigr) \delta_{pr}  \nn
		&\quad + e^2 \Bigl( \frac{32}{3} \q_e \q_d - 24 \q_d^2 \Bigr) \lwc{e \gamma}{}[][pw] \lwc{e \gamma}{*}[][rw] \delta_{st}  + e^2 \Bigl( \frac{32}{3} \q_e \q_d - 24 \q_e^2 \Bigr)  \lwc{d \gamma}{*}[][ws] \lwc{d \gamma}{}[][wt] \delta_{pr}  \nn
		&\quad + 2 e^2 \q_d \q_e \zeta_e   \delta_{pr} \delta_{st}  \, , \\
\nnn
	\dlwc{eu}{RR}[V][prst] &= \frac{4}{3} e^2 \q_e \delta_{pr} \begin{aligned}[t]
			& \bigg[ N_c \q_d \Bigl( \lwc{du}{LR}[V][wwst]+\lwc{ud}{RR}[V][stww] ) + N_c \q_u \Bigl( \lwc{uu}{LR}[V][wwst]+2\lwc{uu}{RR}[V][stww] \Bigr) \nn
			& +\q_e \Bigl( \lwc{eu}{LR}[V][wwst]+\lwc{eu}{RR}[V][wwst] \Bigr) + 2 \q_u \lwc{uu}{RR}[V][swwt] \bigg] \end{aligned} \nn
		&\quad + \frac{4}{3} e^2 \q_u \delta_{st} \begin{aligned}[t]
			& \bigg[ N_c \q_d \Bigl( \lwc{de}{LR}[V][wwpr]+\lwc{ed}{RR}[V][prww] \Bigr) + N_c \q_u \Bigl( \lwc{ue}{LR}[V][wwpr]+\lwc{eu}{RR}[V][prww] \Bigr) \nn
			& +\q_e \Bigl( \lwc{ee}{LR}[V][wwpr]+4\lwc{ee}{RR}[V][prww] \Bigr) \bigg] \end{aligned} \nn
		&\quad + 12 e^2 \q_e \q_u \lwc{eu}{RR}[V][prst]  - \frac{16}{3} C_F e^2 \q_e \q_u \lwc{uG}{*}[][ws] \lwc{uG}{}[][wt] \delta_{pr} + 8 C_F e g \q_e \Bigl( \lwc{uG}{*}[][ws] \lwc{u \gamma}{}[][wt] + \lwc{u \gamma}{*}[][ws] \lwc{uG}{}[][wt] \Bigr) \delta_{pr} \nn
		&\quad + e^2 \Bigl( 24 \q_u^2 + \frac{32}{3} \q_e \q_u \Bigr) \lwc{e \gamma}{*}[][wp] \lwc{e \gamma}{}[][wr] \delta_{st}   + e^2 \Bigl( 24 \q_e^2 + \frac{32}{3} \q_e \q_u \Bigr)  \lwc{u \gamma}{*}[][ws] \lwc{u \gamma}{}[][wt] \delta_{pr}  \nn
		&\quad + 2 e^2 \q_e \q_u   \zeta_e  \delta_{pr} \delta_{st}  \, ,
\nnn
\eea
\bea
	\dlwc{ed}{RR}[V][prst] &= \frac{4}{3} e^2 \q_e \delta_{pr} \begin{aligned}[t]
			& \bigg[ N_c \q_d \Bigl( \lwc{dd}{LR}[V][wwst]+2\lwc{dd}{RR}[V][stww] \Bigr) + N_c \q_u \Bigl( \lwc{ud}{LR}[V][wwst]+\lwc{ud}{RR}[V][wwst] \Bigr) \nn
			&+ \q_e \Bigl(\lwc{ed}{LR}[V][wwst]+\lwc{ed}{RR}[V][wwst]\Bigr) + 2 \q_d \lwc{dd}{RR}[V][swwt] \bigg] \end{aligned} \nn
		&\quad + \frac{4}{3} e^2 \q_d \delta_{st} \begin{aligned}[t]
			& \bigg[ N_c \q_d \Bigl( \lwc{de}{LR}[V][wwpr]+\lwc{ed}{RR}[V][prww] \Bigr) + N_c \q_u \Bigl( \lwc{ue}{LR}[V][wwpr]+\lwc{eu}{RR}[V][prww] \Bigr) \nn
			& + \q_e \Bigl( \lwc{ee}{LR}[V][wwpr]+4\lwc{ee}{RR}[V][prww] \Bigr) \bigg] \end{aligned} \nn
		&\quad + 12 e^2 \q_d \q_e \lwc{ed}{RR}[V][prst]   - \frac{16}{3} C_F e^2 \q_e \q_d \lwc{dG}{*}[][ws] \lwc{dG}{}[][wt] \delta_{pr}   + 8 C_F e g \q_e \Bigl( \lwc{dG}{*}[][ws] \lwc{d \gamma}{}[][wt] + \lwc{d \gamma}{*}[][ws] \lwc{dG}{}[][wt] \Bigr) \delta_{pr}  \nn
		&\quad + e^2 \Bigl( 24 \q_d^2 + \frac{32}{3} \q_e \q_d \Bigr) \lwc{e \gamma}{*}[][wp] \lwc{e \gamma}{}[][wr] \delta_{st}  + e^2 \Bigl( 24 \q_e^2 + \frac{32}{3} \q_e \q_d \Bigr)  \lwc{d \gamma}{*}[][ws] \lwc{d \gamma}{}[][wt] \delta_{pr} \nn
		&\quad + 2 e^2 \q_e \q_d   \zeta_e  \delta_{pr} \delta_{st}   \, , \\
\nnn
	\dlwc{ue}{LR}[V][prst] &= \frac{4}{3} e^2 \q_u \delta_{pr} \begin{aligned}[t]
			& \bigg[ N_c \q_d \Bigl(\lwc{de}{LR}[V][wwst]+\lwc{ed}{RR}[V][stww]\Bigr) + N_c \q_u \Bigl(\lwc{ue}{LR}[V][wwst]+\lwc{eu}{RR}[V][stww] \Bigr) \nn
			& + \q_e \Bigl(\lwc{ee}{LR}[V][wwst]+4\lwc{ee}{RR}[V][stww]\Bigr) \bigg] \end{aligned} \nn
		&\quad + \frac{4}{3} e^2 \q_e \delta_{st} \begin{aligned}[t]
			& \bigg[ N_c \q_d \Bigl(\lwc{ud}{LL}[V1][prww]+\lwc{ud}{LR}[V1][prww]\Bigr) + N_c \q_u \Bigl(2\lwc{uu}{LL}[V][prww]+\lwc{uu}{LR}[V1][prww] \Bigr) \nn
			& + \q_e \Bigl(\lwc{eu}{LL}[V][wwpr]+\lwc{ue}{LR}[V][prww]\Bigr) + 2 \q_u \lwc{uu}{LL}[V][pwwr] \bigg] \end{aligned} \nn
		&\quad - 12 e^2 \q_e \q_u \lwc{ue}{LR}[V][prst]  - \frac{16}{3} C_F e^2 \q_e \q_u \lwc{uG}{}[][pw] \lwc{uG}{*}[][rw] \delta_{st}  + 8 C_F e g \q_e \Bigl( \lwc{uG}{}[][pw] \lwc{u\gamma}{*}[][rw] + \lwc{u\gamma}{}[][pw] \lwc{uG}{*}[][rw] \Bigr) \delta_{st}  \nn
		&\quad + e^2 \Bigl( \frac{32}{3} \q_e \q_u - 24 \q_e^2 \Bigr)  \lwc{u \gamma}{}[][pw] \lwc{u \gamma}{*}[][rw] \delta_{st} + e^2 \Bigl( \frac{32}{3} \q_e \q_u - 24 \q_u^2 \Bigr)  \lwc{e \gamma}{*}[][ws] \lwc{e \gamma}{}[][wt]  \delta_{pr}  \nn
		&\quad + 2 e^2 \q_u \q_e \zeta_e   \delta_{pr} \delta_{st}  \, , \\
\nnn
	\dlwc{de}{LR}[V][prst] &= \frac{4}{3} e^2 \q_d \delta_{pr} \begin{aligned}[t]
			& \bigg[ N_c \q_d \Bigl(\lwc{de}{LR}[V][wwst]+\lwc{ed}{RR}[V][stww]\Bigr) + N_c \q_u \Bigl(\lwc{ue}{LR}[V][wwst]+\lwc{eu}{RR}[V][stww] \Bigr) \nn
			& + \q_e \Bigl(\lwc{ee}{LR}[V][wwst]+4\lwc{ee}{RR}[V][stww]\Bigr) \bigg] \end{aligned} \nn
		&\quad + \frac{4}{3} e^2 \q_e \delta_{st} \begin{aligned}[t]
			& \bigg[ N_c \q_d \Bigl(2\lwc{dd}{LL}[V][prww]+\lwc{dd}{LR}[V1][prww]\Bigr) + N_c \q_u \Bigl(\lwc{ud}{LL}[V1][wwpr]+\lwc{du}{LR}[V1][prww]\Bigr) \nn
			& + \q_e \Bigl(\lwc{ed}{LL}[V][wwpr]+\lwc{de}{LR}[V][prww]\Bigr) + 2 \q_d \lwc{dd}{LL}[V][pwwr] \bigg] \end{aligned} \nn
		&\quad -12 e^2 \q_d \q_e \lwc{de}{LR}[V][prst]   - \frac{16}{3} C_F e^2 \q_e \q_d \lwc{dG}{}[][pw] \lwc{dG}{*}[][rw] \delta_{st}  + 8 C_F e g \q_e \Bigl( \lwc{dG}{}[][pw] \lwc{d\gamma}{*}[][rw] + \lwc{d\gamma}{}[][pw] \lwc{dG}{*}[][rw] \Bigr) \delta_{st}  \nn
		&\quad + e^2 \Bigl( \frac{32}{3} \q_e \q_d - 24 \q_e^2 \Bigr) \lwc{d \gamma}{}[][pw] \lwc{d \gamma}{*}[][rw] \delta_{st}  + e^2 \Bigl( \frac{32}{3} \q_e \q_d - 24 \q_d^2 \Bigr) \lwc{e \gamma}{*}[][ws] \lwc{e \gamma}{}[][wt]  \delta_{pr} \nn
		&\quad + 2 e^2 \q_d \q_e \zeta_e   \delta_{pr} \delta_{st}  \, , 
\nnn
\eea
\subsection{SMEFT matching}
\label{sec:SMEFT}

\begin{table}[h!]
\label{tab:LEFTSMEFTscalar}
    \centering
\begin{tabular}{|c|c|}
\hline
Scalar Operators &  Matching \\
\midrule\midrule
$\op{eu}{S}{RR}$  &  $-C^{(1)}_{\substack{ lequ \\ prst}}$
 \\
$\op{ed}{S}{RR}$ &  0
 \\
$\op{eu}{S}{RL} $  & 0
 \\
$\op{ed}{S}{RL}$  & $C_{\substack{ ledq \\ prst}}$ \\
\bottomrule
\end{tabular}
\caption{LEFT-SMEFT matching of Scalar operators at tree level~\cite{Jenkins:2017jig}. }
\qquad
\begin{tabular}{|c|c|}
\hline
Vector Operators &  Matching \\
\midrule\midrule
$\op{eu}{V}{LL}$ & $C^{(1)}_{\substack{ lq \\ prst}} -  C^{(3)}_{\substack{ lq \\ prst}}
-\frac{\gcZ^2}{ M_Z^2}   \left[Z_{e_L} \right]_{pr} \left[Z_{u_L} \right]_{st} $ \\
$\op{ed}{V}{LL}$  & $C^{(1)}_{\substack{ lq \\ prst}} + C^{(3)}_{\substack{ lq \\ prst}} 
-\frac{\gcZ^2}{ M_Z^2}   \left[Z_{e_L} \right]_{pr} \left[Z_{d_L} \right]_{st}$  \\
\hline 
$\op{eu}{V}{RR}$ & $C_{\substack{ eu \\ prst}} -\frac{\gcZ^2}{ M_Z^2}   \left[Z_{e_R} \right]_{pr} \left[Z_{u_R} \right]_{st} $\\
$\op{ed}{V}{RR}$  &   $C_{\substack{ ed \\ prst}} -\frac{\gcZ^2}{M_Z^2}   \left[Z_{e_R} \right]_{pr} \left[Z_{d_R} \right]_{st}$ \\
\hline
$\op{eu}{V}{LR} $ & $C_{\substack{ lu \\ prst}} 
-\frac{\gcZ^2}{M_Z^2}   \left[Z_{e_L} \right]_{pr} \left[Z_{u_R} \right]_{st}$ \\
$\op{ed}{V}{LR}$  &$ C_{\substack{ ld \\ prst}} 
-\frac{\gcZ^2}{M_Z^2}   \left[Z_{e_L} \right]_{pr} \left[Z_{d_R} \right]_{st} $\\
$\op{eu}{V}{RL}$ & $C_{\substack{ qe \\ prst}}
 -\frac{\gcZ^2}{M_Z^2}   \left[Z_{u_L} \right]_{pr} \left[Z_{e_R} \right]_{st}$  \\
$\op{ed}{V}{RL}$ & $ C_{\substack{ qe \\ prst}} 
-\frac{\gcZ^2}{ M_Z^2}    \left[Z_{d_L} \right]_{pr} \left[Z_{e_R} \right]_{st} $ \\
\bottomrule
\end{tabular}
\caption{LEFT-SMEFT matching of Vector operators at tree level~\cite{Jenkins:2017jig}. }
\end{table}

where
\bea
[Z_{e_L}]_{pr} &= \left[\delta_{pr}\left(-\frac 12+\sc^2 \right) - \frac12 v_T^2  C^{(1)}_{\substack {Hl \\  pr}} - \frac12 v_T^2  C^{(3)}_{\substack {Hl \\  pr}} \right], \quad
[Z_{e_R}]_{pr} &= \left[\delta_{pr}\left(+\sc^2 \right) - \frac12 v_T^2  C_{\substack {He \\  pr}}  \right], 
\eea
\bea
[Z_{u_L}]_{pr} &=  \left[\delta_{pr}\left(\frac 12-\frac 23 \sc^2 \right) - \frac12 v_T^2  C^{(1)}_{\substack {Hq \\  pr}} + \frac12 v_T^2  C^{(3)}_{\substack {Hq \\  pr}} \right],  \quad
[Z_{u_R}]_{pr} &=  \left[\delta_{pr}\left(-\frac 23 \sc^2 \right) - \frac12 v_T^2  C_{\substack {Hu \\  pr}}  \right], 
\eea
\bea
[Z_{d_L}]_{pr} &=  \left[\delta_{pr}\left(-\frac 12+ \frac 13 \sc^2 \right) - \frac12 v_T^2  C^{(1)}_{\substack {Hq \\  pr}} - \frac12 v_T^2  C^{(3)}_{\substack {Hq \\  pr}}  \right], \quad
[Z_{d_R}]_{pr} &=  \left[\delta_{pr}\left(+\frac13\sc^2 \right) - \frac12 v_T^2  C_{\substack {Hd \\  pr}}  \right] .
\eea

\bibliographystyle{ieeetr}
\bibliography{Pi0mue}

\end{document}